# Magnetic properties of a highly ordered single crystal of the layered perovskite YBaCuFe$_{0.95}$Mn$_{0.05}$O$_5$

Xiaodong Zhang[1*], Arnau Romaguera[1], Felip Sandiumenge[1], Oscar Fabelo[2], Javier Blasco[3], Javier Herrero-Martín[4] and José Luis García-Muñoz[1]

[1]*Institut de Ciència de Materials de Barcelona, ICMAB-CSIC, Campus UAB, 08193 Bellaterra, Spain*
[2]*ILL-Institut Laue Langevin, 38042 Grenoble Cedex, France.*
[3]*Instituto de Nanociencia y Materiales de Aragón, Departamento de Física de la Materia Condensada, CSIC-Universidad de Zaragoza, 50009 Zaragoza, Spain*
[4]*CELLS-ALBA Synchrotron, 08290 Cerdanyola del Vallès, Barcelona, Spain.*

**Abstract**

The layered perovskite YBaCuFeO$_5$ (YBCFO) is able to adopt chiral magnetic order up to unexpectedly high temperatures, paving the way to strong magnetoelectric coupling at room temperature. In this perovskite A-site cations are fully ordered whereas the occupancy of the B-sites strongly depend on the preparation process. Though the structure is not geometrically frustrated, the presence of partial Fe$^{3+}$/Cu$^{2+}$ disorder at the B-sites produces magnetic frustration. In an effort to increase the spin-orbit coupling in the system, we have synthesized and studied YBaCuFe$_{0.95}$Mn$_{0.05}$O$_5$ in single crystal form, where the highly symmetric Fe$^{3+}$ ions (3d$^5$) are partially substituted with Jahn-Teller active 3d$^4$ Mn$^{3+}$ ions. We report the structural and magnetic properties of a highly ordered single crystal of this layered perovskite, which are presented in comparison with a polycrystalline specimen (three times more disordered). Single-crystal neutron diffraction measurements reveal two collinear magnetic phases and the lack of incommensurate spiral order. The magnetic phases and transitions found in the crystal grown by the traveling solvent floating zone (TSFZ) method are fully described and analyzed in the light of its high level of Fe/Cu cationic order (~90%).

**\* Corresponding autor:**
Xiaodong Zhang
E-mail:xzhang@icmab.es
*Institut de Ciència de Materials de Barcelona, ICMAB-CSIC, Campus UAB, 08193 Bellaterra, Spain*

### 1. Introduction

Among the novel properties realized in chiral magnetic systems the multiferroicity is one of the most attractive. Most of the chiral magnetoelectric multiferroics investigated in recent years are geometrically frustrated magnets where the spin-orbit coupling favors noncollinear (spiral) magnetic ordering through the Dzyaloshinskii-Moriya (DM) mechanism. Unfortunately, magnetic ordering temperatures are too low (typically $T_S<100$ K for spiral orders [1,2]). Exceptionally the layered perovskite $YBaCuFeO_5$ (YBCFO) has been reported to display magnetism-driven ferroelectricity at unexpectedly high temperatures [3–7]. Its structure exhibits alternation of $A^{3+}/A^{2+}$ layers along the *c*-axis ($A^{3+}=Y^{3+}$ or $R^{3+}$ [rare earth] and $A^{2+}=Ba^{2+}$). B-site metals occupy $CuFeO_9$ bilayers of corner-sharing $Cu^{2+}O_5$ and $Fe^{3+}O_5$ square pyramids parallel to the *ab* plane, separated by $Y^{3+}$ (or $R^{3+}$) layers. The $Ba^{2+}$ ions are located within the bilayer spacing. As regards its magnetic transitions, upon cooling a commensurate (CM) $\mathbf{k}_1=(1/2,1/2,1/2)$ antiferromagnetic phase below $T_{N1}= 440K$ is followed by an incommensurate (ICM) spiral magnetic order ($\mathbf{k}_2 = (1/2,1/2,1/2 \pm q)$,) at $T_{N2}$ (ranging between 150 and 310 K [7]) that persists down to the base temperature. Though YBCFO is not a geometrically frustrated magnet, magnetic frustration gets present due to partial Fe/Cu disorder in the structure. As a consequence, YBCFO displays an extraordinary tunability of its incommensurate (spiral) ordering temperature ($T_S=T_{N2}$) by manipulating the Cu/Fe chemical disorder in the bipyramids by means of the synthesis procedure [7]. Likewise, beyond cation disorder, the strength of the main magnetic interactions can be also tuned by chemical pressure [8]. The exchange coupling $J_{c2}$ between a Cu-Fe pair in a bipyramid is the only FM interaction, whereas magnetic coupling between successive bipyramids along-*c* ($J_{c1}$) is antiferromagnetic (AFM). Magnetic exchange between neighbor bipyramids within the *ab* plane is always AFM, independently of the Cu/Fe disorder. This explains the very high Néel temperature $T_{N1}\approx 440K$ of the CM collinear AFM phase ($\mathbf{k}_1=(½, ½, ½)$). S

In the presence of Cu/Fe disorder, $J_{c2}$ (within a bipyramid) can be of different sign and magnitude: (i) a strong AFM interaction between two Fe ions; (ii) a FM exchange in Fe-Cu pairs or (iii) a weak AFM Cu-Cu exchange [7]. The model developed by Scaramucci *et al*. [9,10] justifies the onset of a ICM spiral phase ($\mathbf{k}_2=(1/2,1/2,1/2\pm q)$ below $T_{N2}<T_{N1}$. Essential ingredients for this mechanism based on Heisenberg spins with only NN interactions are: (i) Cu/Fe disorder permitting the presence of Fe/Fe bipyramids with strong AFM $Fe^{3+}$-O-$Fe^{3+}$ bonds, always parallel to *c*; (ii) the occurrence of local canting of the spins associated to these Fe/Fe bonds and (iii) a long range coupling between local cantings favoured by neighbouring Fe/Fe impurity bonds in the *ab* layers. The relevance of the spin-orbit coupling in this structure is not clear yet. The similarity of $Mn^{3+}$ and $Fe^{3+}$ ionic sizes (both ≈0.58 Å in pyramidal coordination [11]) allows to explore the possibility of increasing the spin-orbit coupling in $YBaCu(Fe_{1-x}Mn_x)O_5$ through partial substitution of highly symmetric $Fe^{3+}$ ($3d^5$) with less symmetric $Mn^{3+}$ ions ($3d^4$, Jahn-Teller active) [12]. The absence of ferroelectricity in the YBCFO crystal of Ref. [13] was attributed to a spiral rotation plane perpendicular to the c-axis. Doping with Mn the Fe pyramids has positive effects on the orientation of the spiral for the activation of the DM mechanism, as reported recently in polycrystalline $YBaCu(Fe_{1-x}Mn_x)O_5$ samples [12]. The presence of Mn produces a progressive reorientation of the tilt $\theta_{spiral}$ of the rotation plane of the spins in the spiral phase. It moves away from the *ab* plane where the DM-based models predict null spontaneous polarization.

In this work we report the magnetic properties of a highly ordered single crystal of the layered perovskite $YBaCuFe_{0.95}Mn_{0.05}O_5$ fabricated following the traveling solvent floating zone (TSFZ) method. Its properties are discussed in comparison with those of a polycrystalline sample of the same composition that presents less cationic order.

## 2. Experimental

Polycrystalline $YBaCuFe_{0.95}Mn_{0.05}O_5$ (hereafter denoted as "powder") was prepared through the conventional solid-state reaction method. Stoichiometric amounts of $Y_2O_3$, $BaCO_3$, CuO, $Fe_2O_3$ and $Mn_2O_3$ were dried and weighed. A pre-annealing process was performed on the $Y_2O_3$ oxide at 900 °C for 10 h. All precursors were thoroughly mixed, grounded and pressed into pellets, and then sintered in a tubular furnace at 1100 °C for 50 h in air. Finally, the sample was cooled down to room temperature (RT) inside the furnace at a controlled rate of 300 °C/h. The sample quality was assessed by X-ray powder diffraction using a Siemens D-5000 diffractometer ($\lambda[Cu\ K_\alpha]$=1.54 Å). A part of the resulting material was also used to prepare the polycrystalline feed rod for the growth of a $YBaCuFe_{0.95}Mn_{0.05}O_5$ single-crystal (hereafter denoted as "single-crystal"). The powder was loaded and subsequently packed with cylindrical shape and then compacted under 1500 bar of isostatic pressure. The solid cylindrical rod was further compacted by sintering at 1100 °C for 50 h in air. The final dimensions of the rod were 4 mm in diameter and about 60 mm in length. A solvent consisting of CuO with 2 wt% $B_2O_3$ was used for the crystal growth of the incongruently melting compound [14] using the TSFZ method. This was performed using a four-mirror optical furnace. The two rods were rotated at 25 rpm in opposite directions. Once adjusted the pulling rate, the composition of the solvent reached an equilibrium after several hours of growth, and crystals with the desired YBaCuFe$_{0.95}$Mn$_{0.05}$O$_5$ composition started to precipitate. After several days of stable growth a large high-quality single crystal was obtained.

The magnetic susceptibility of the powder and single-crystal samples was measured using a VSM magnetometer in a Physical Properties Measurement System (PPMS, Quantum Design Inc) under a magnetic field of 2 kOe. In the case of the single-crystal, the field was applied parallel to the *ab*-plane. High resolution transmission electron microscopy images (HRTEM) were obtained at 200kV using a field-emission gun FEI Tecnai F20 S/TEM electron microscope. Samples for HRTEM observation were obtained by gentle mechanical grinding of single crystals. Multislice image simulations were carried out with the EMS software package [15]. The structural analysis of the polycrystalline sample was performed on the basis of synchrotron X-ray diffraction (SXRD) data collected at room temperature (RT), in the MSPD beamline of the ALBA Synchrotron (Barcelona, Spain), using λ=0.41338 Å. Neutron diffraction experiments were performed on both samples at the high-flux reactor of the Institut Laue Langevin (ILL, Grenoble, France). Neutron powder diffraction (NPD) patterns on the powder sample were collected using the high intensity diffractometer D1B (λ=2.52 Å) between 10 and 500 K. The homogeneity and orientation of the single-crystal was assessed using Orient Express. Further temperature dependent reciprocal space maps were recorded using the Cyclops Laue diffractometer between 50 and 300 K. Finally, additional measurements were carried out at the D9 four-circle neutron diffractometer (λ=0.836 Å), in which a large set of 833 nuclear Bragg reflections (456 independent) were collected at 50 K to obtain a good structural model that allowed us to determine the Fe/Cu cationic order and interatomic distances. A small two-dimensional (2D) area detector of 6 × 6 cm (32 × 32 pixels) allows reciprocal space survey and optimization of the peak position. The program RACER [16] was used to integrate the omega- and omega-2theta-scans and to correct them for the Lorentz factor. q-scans were collected along specific directions to assess the presence of the different propagation vectors. The temperature dependence of particular peaks was monitored between 50 K and RT using a displex cryostat. Magnetic data collections were made at 50 K. Structural and magnetic refinements were carried out using the FullProf set of programs [17]. Neutron refinements were done by least square minimization of the integrated intensities and extinction corrections were applied following the model of Becker-Coppens [18].

## 3. Results and discussion

### *3.1 Structural characterization*

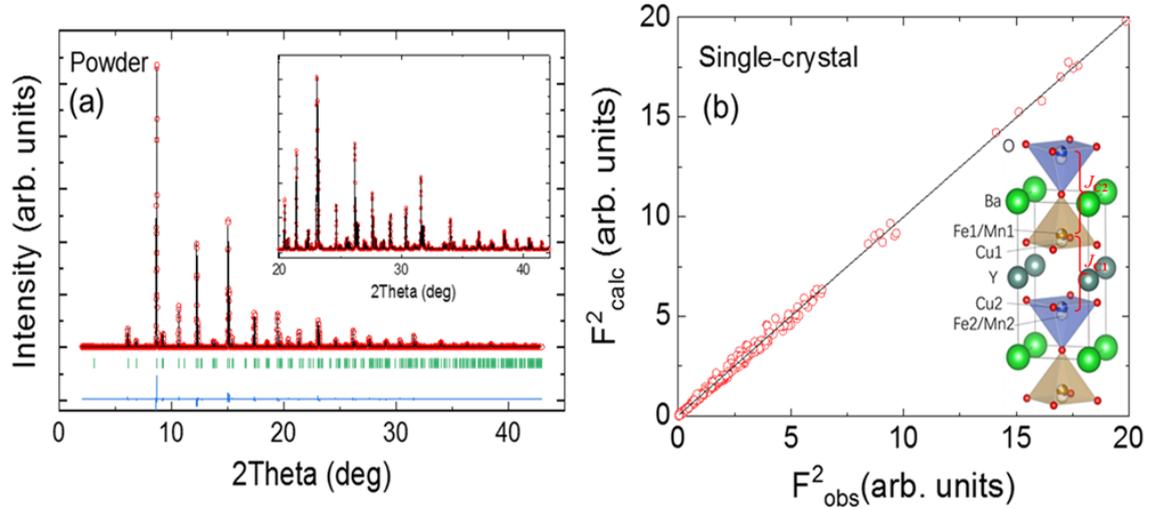

**Fig. 1.** (a) [**powder** sample] Rietveld refinement (black curve) of the synchrotron X-ray diffracted intensities (red circles) at 300 K. The bottom blue line is the observed-calculated difference. Inset: expanded region at high-angles; (b) [**crystal**] Agreement plot of the refinement on nuclear reflections of the single-crystal collected at 50 K ($\lambda$=0.836 Å). The calculated structure factors are plotted against the experimental ones. Inset: *P4mm* structure showing the corner-sharing square bipyramidal layers. The color of each pyramid corresponds to that of the dominant cation in it (blue: Cu; brown: Fe).

Fig. 1(a) shows the refined SXRD pattern (at 300 K) for the powder sample ($\chi^2$: 70.9, $R_B$: 5.44, $R_f$: 7.16), and Fig. 1(b) the neutron integrated intensity refinement (at 50 K) for the single-crystal ($\chi^2$: 1.12, $R_F$: 2.95, $R_F^2{}_W$: 4.23). The atomic positions and the occupancies of the Cu and Fe[Mn] ions in the two pyramids of the unit cell were refined using the *P4mm* symmetry. The z-coordinates of the same metal in upper and lower positions were constrained by z(M1)+z(M2)=1. The positions of the symmetry-inequivalent oxygen atoms were refined independently in both samples. The detailed structural information obtained for both specimens is summarized in Table I. A key feature with strong influence on the magnetic properties is the Fe/Cu chemical disorder. For that reason, the partial occupation by the Cu and Fe[Mn] of the upper and lower pyramids in the structure (see inset of Fig. 1(b)) was carefully refined. The results about the chemical disorder in the two samples with identical composition are given in Table I. The proper occupancy (Occ) in the pyramids was 0.70(2) for the powder sample and 0.89(3) for the single-crystal.

Remarkably, this implies that ≈30% of nominal Fe pyramids are occupied by Cu in the powder sample, whereas in the single-crystal this is so only in ≈10% of the pyramids (an occupation Occ=0.5 would correspond to a random B-site cation distribution, and an Occ=1 describes a fully ordered Fe/Cu structure). Therefore, the powder sample is much more disordered than the single-crystal. Next, we present a comparison of the magnetic properties in both samples.

**Table I:** Structural parameters obtained from SXRD data (MSPD@Alba) at 300 K (*powder*) and single-crystal neutron diffraction data (D9@ILL) at 50 K (**single-crystal**), respectively. Ba is at the origin. The "Occ" or "Chemical order" refers to the chemical occupation of the dominant cation in its square pyramid (see inset of Fig. 1(b)). The dominant cation in upper pyramid (brown) and lower pyramid (blue) being Fe and Cu, respectively.

| sample | P4mm | z (Y) (0 0 z) | z (Fe1) (1/2 1/2 z) | z(Cu2) (1/2 1/2 z) | z(O$_1$) | z(O$_2$) (0 1/2 z) | z(O$_3$) (0 1/2 z) | Occ (Chemical order) |
|---|---|---|---|---|---|---|---|---|
| Powder | a=3.8735(3) Å c=7.6648(3) Å | 0.5116(4) | 0.7551(6) | 0.2850(3) | 0.011(3) | 0.327(1) | 0.697(1) | 0.704(22) |
| Single-crystal | a=3.8699(3) Å c=7.6359(3) Å | 0.4979(3) | 0.7343(8) | 0.2674(8) | -0.002(5) | 0.312 (4) | 0.682(4) | 0.891(26) |

Fig. 2(a) shows a HRTEM image of the single-crystal sample taken along the [100] zone axis. Image simulations for two different thicknesses (4.6 and 1.5 nm) at a defocus value of 68 nm are over-imposed as insets **1** and **2**, respectively, where atom columns are seen dark. Still in Fig. 2(a), the upper right inset shows the atomic positions on the simulated image. The prominent bright fringes correspond to the Y planes of the crystal structure, as indicated by a yellow arrow. It can be observed that, in the thicker region, the contrast above and below these planes is not symmetric: the atomic planes containing the Cu ions are brighter than those containing the Fe(Mn) ones, giving rise to a characteristic double fringe contrast. This contrast effect, on the other hand, disappears in the thinner area close to the edge of the crystallite at the bottom of the image. It is interesting to note that image simulations using crystal data from a non-Mn doped sample revealed symmetric contrast about the prominent fringes. Therefore, the double fringe contrast can be attributed to the differential distortion of the Cu and Fe sites induced by the selective Mn doping. A more careful inspection reveals lateral variations of the Cu-plane intensity in the thicker area which, on the basis of the above considerations, can be interpreted as variations in the level of lattice distortion induced by concomitant variations of the Mn concentration.

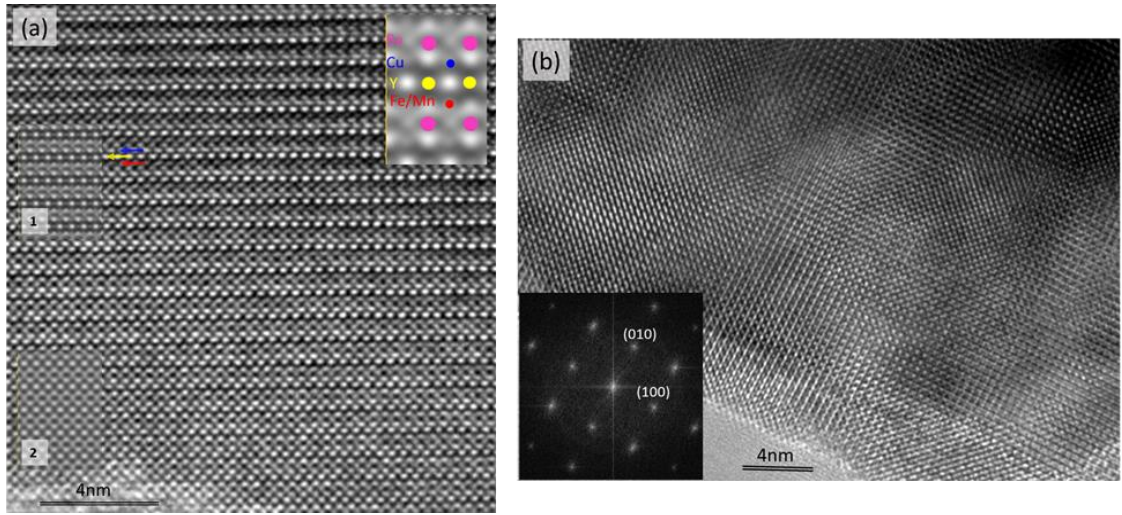

**Fig. 2.** YBaCuFe$_{0.95}$Mn$_{0.05}$O$_5$. (a) HRTEM image of the single-crystal viewed along the [100] zone axis. Insets labelled **1** and **2** are simulations for thickness values of 4.6 and 1.5 nm, respectively (defocus 68 nm). The inset at the upper right corner shows the position of atomic columns. (b) HRTEM image of the same sample viewed along the [001] zone axis. The inset shows the corresponding fast Fourier transform.

The HRTEM image shown Fig. 2(b) corresponds to the basal plane of the structure (zone axis [001]). The most apparent feature of the image is a patched like contrast defining very small domains, 4-6 nm in size. This nanostructure was observed in all crystallites studied in this orientation. Dark contrasts in these HRTEM images indicate inhomogeneous strains, while according to image simulations the elongated shape of the bright dots observed, for instance in the right side of the image, correspond to <110> tilts of the lattice away from the zone axis. The fast Fourier transform (see inset) shows diffuse scattering effects around the spots which can be associated to local disorder and strain inhomogeneities observed in the image. It is worth mentioning that this patched nanostructure was not observed in images obtained from non-doped samples.

*3.2 Magnetic characterization and neutron diffraction*

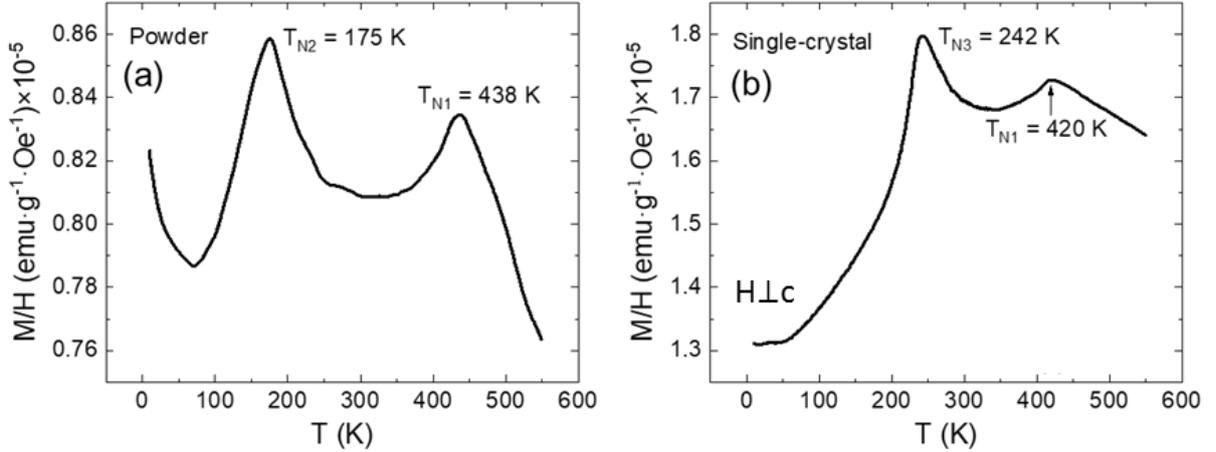

**Fig. 3.** Magnetic susceptibility curves (2 kOe) for (a) the powder and (b) the single-crystal samples. Transition temperatures are indicated in the curves.

Fig. 3 displays the magnetic susceptibility for the two studied samples (in field-cooling conditions, FC) from 10 to 550 K. Two magnetic transitions can be observed in both samples. The highest transition, corresponding to the collinear $\mathbf{k_1}$=(½, ½, ½) phase peaks at $T_{N1}$=438 K (powder) and 420 K (crystal). The lowest transition temperature deviates around 70 K comparing both samples but, as we are showing later, the low-temperature transition corresponds to different spin configurations in our powder and single crystal samples.

Temperature-dependent neutron diffraction measurements were carried out on the powder sample in the 10 to 500 K range. Fig. 4(a) plots a Q-T projection of neutron diffracted intensities around the main (1/2 1/2 1/2) magnetic reflection. In agreement with the magnetic susceptibility, the two transitions correspond to (i) the CM phase which appears at the onset temperature $T_{N1}$, with propagation vector $\mathbf{k_1}$=(1/2, 1/2, 1/2), and (ii) the ICM phase, showing up at $T_{N2}$ (=$T_S$, the spiral transition) with characteristic split satellite reflections and propagation vector $\mathbf{k_2}$=(1/2, 1/2, 1/2±q). An ICM modulation q=0.102 r.l.u. was found at the lowest temperature. Fig. 4(b) illustrates the neutron powder diffraction pattern collected at 50 K on D1B that shows the dominant ICM magnetic reflections from the spiral order appearing in the low-angle region. A small residual amount of the high-temperature CM collinear phase is observed. The temperature dependence of the averaged ordered magnetic moments for the CM and ICM phases against temperature is depicted in Fig.4(c).

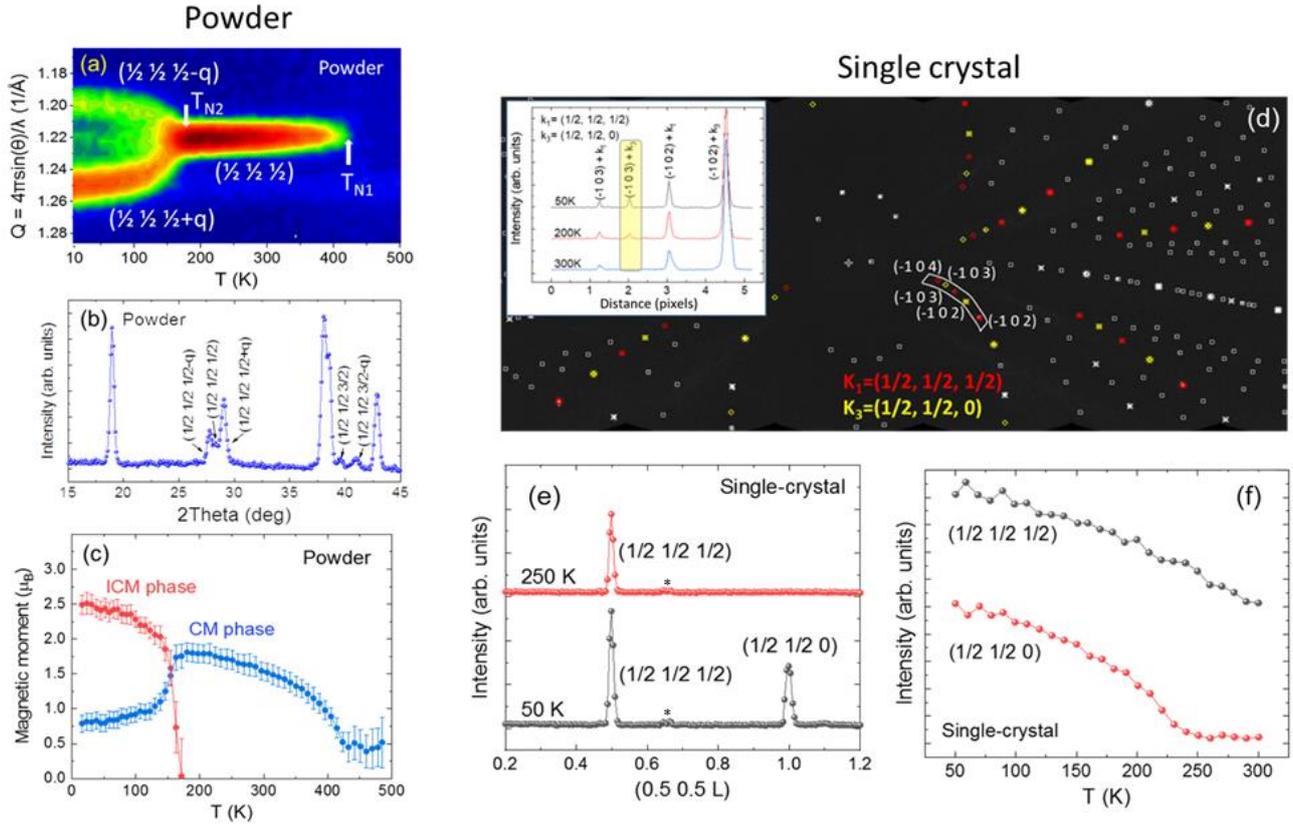

**Fig. 4. Powder.** (a) Q-T projection of the temperature dependence for the neutron-diffracted intensities around the (½ ½ ½) reflection for the powder sample. (b) Low-angle region of the neutron diffraction pattern recorded at 50 K for the powder sample; (c) T-dependence of the (average) ordered magnetic moments associated to the CM and ICM phases in the powder sample (adapted from our ref. [12]). **Crystal.** (d) Section of a Laue neutron diffraction pattern taken at 50K, showing the indexation of nuclear reflections (in white) and the magnetic ones with $\mathbf{k}_1$=[1/2, 1/2, 1/2] and $\mathbf{k}_3$=[1/2, 1/2, 0] propagation vectors (in red and yellow, respectively; some magnetic reflections have also nuclear contribution). (inset) Comparison of the integrated profiles (along the curve indicated in the Laue) at three different temperatures. (e) $Q_L$-scans in the single-crystal centered along (0.5, 0.5, L) line (range 0.2<L<1.2). Comparison of the scans recorded at 50 K and 250 K. (f) Temperature dependence of the neutron integrated intensities of (½ ½ ½) [$\mathbf{k_1}$] and (½ ½ 0) [$\mathbf{k_2}$] magnetic reflections.

The magnetic behaviour is however remarkably different for the single-crystal. Temperature-dependent neutron diffraction measurements performed using the Cyclops Laue diffractometer at ILL revealed a non-conventional magnetic behaviour in the crystal, qualitatively different to the polycrystalline sample of the same composition. Interestingly, the indexation of the magnetic reflections at 50K (Fig. 4d) discloses that the ICM satellites characteristic of the spiral phase have been suppressed and replaced by a new translational symmetry with $\mathbf{k}_3$=(1/2, 1/2, 0) magnetic propagation vector. This was further corroborated by $Q_L$-scans performed at 50 K and 250 K on D9 along the (0.5, 0.5, L) line (in the range 0.2<L<1.2) with the purpose of tracing the possible magnetic reflections, as shown in Fig. 4(e). The observation of (0.5 0.5 0.5) and (0.5 0.5 0) magnetic reflections at 50 K confirms the coexistence of respectively $\mathbf{k}_1$ and $\mathbf{k}_3$ magnetic phases below $T_{N3}$, whereas at 250 K only $\mathbf{k}_1$ reflections are observed. This is also evident from the evolution of the neutron-diffracted intensities of the two magnetic reflections versus temperature as shown in Fig. 4(f). Upon warming from 50 to 300 K, the unexpected AF3 phase ($\mathbf{k}_3$) disappears around 240 K, while AF1 ($\mathbf{k}_1$) keeps being present up to much higher temperatures (Fig. 4f). The observations are in agreement with the transitions exposed in the susceptibility at $T_{N3}$ = 242 K and $T_{N1}$ = 420 K, corresponding to the ordering of AF3 and AF1 phases, respectively. The magnetic evolution disclosed in Fig. 4(f) suggests that the two magnetic phases very likely develop in distinct regions of the crystal. In Fig. 4(e) (see the asterisk) a tiny protuberance was observed around (0.5 0.5 0.65) which probably comes from a tiny misaligned domain. A main result of the neutron study was therefore the lack of the ICM (spiral) magnetic phase in the single crystal.

### *3.3 Magnetic structures.*

No previous reports in the literature deal with the orientation of the moments in the tetragonal *ab* plane for YBCFO type compounds, due to the inherent limitations of the *neutron powder diffraction*. We could overcome these intrinsic limitations by the analysis of the various sets of single-crystal magnetic reflections collected on D9, which enabled us to determine the spin configuration and the orientation of the spins with respect to the tetragonal unit cell for the two coexisting CM phases at 50 K.

In the crystal, collinear order models were used to fit the magnetic reflections associated to each propagation vector. For determining the magnetic structures in the crystal, the crystallographic parameters and the scale factors were fixed to the values obtained in the nuclear refinements. The phase difference between the magnetic moments at the two sites was fixed to 180º as found in earlier refs. [5,13]. The amplitudes of the ordered magnetic moments at the nominally Fe and Cu sites were refined independently, without constraints, but referred to the total volume of the crystal. The spin orientation can be defined by the polar angles $\theta$ and $\varphi$. The former measures the angular distance of the spin to the *c* axis ($\theta$), and $\varphi$ measures the distance of the spin component in the **ab**-plane with respect to the **a**-axis. The magnetic refinements unambiguously converged to the same solution independently of the proposed initial values. The collinear spin configurations at 50 K for AF1 and AF3 phases are summarized in the Table II, corresponding respectively to the propagation vectors $\mathbf{k_1}$= (1/2, 1/2, 1/2) and $\mathbf{k_3}$= (1/2, 1/2, 0). Moreover, the AF1 and AF2 (circular spiral) magnetic structures found in the homologue powder sample are described in the same table for the temperatures 300 K (AF1) and 10 K (AF2). In the spiral phase $\varphi$ is not defined and $\theta$ corresponds to the inclination angle of the spin rotation plane [12].

**Table II**. Magnetic structures: magnetic phases, magnetic moments, spin orientation (in polar coordinates) and agreement factors obtained in the single crystal and in the powder sample. Moments are referred to the total volume of the crystal.

| Crystal | $T_N$ (K) | m(Fe) ($\mu_B$) | m(Cu) ($\mu_B$) | $\theta$ (deg) | $\varphi$ (deg) |
|---|---|---|---|---|---|
| AF1 $\mathbf{k_1}$= (1/2, 1/2, 1/2) | 420 | 2.23 (3) | -0.70 (4) | 70 (2) | -89 (2) |
| 50 K | | $R_F^2$ (%):17.0 | $R_F^2$w (%):18.7 | $R_F$ (%):10.8 | $\chi^2$: 13.7 |
| AF3 $\mathbf{k_3}$= (1/2, 1/2, 0) | 242 | 1.01 (15) | -0.84 (15) | 68 (3) | -26 (3) |
| 50 K | | $R_F^2$ (%):21.3 | $R_F^2$w (%):22.0 | $R_F$ (%):13.5 | $\chi^2$: 15.5 |

| Powder | $T_N$ (K) | m(Fe) ($\mu_B$) | m(Cu) ($\mu_B$) | $\theta$ (deg) | $\varphi$ (deg) |
|---|---|---|---|---|---|
| AF1 $\mathbf{k_1}$= (1/2, 1/2, 1/2) | 438 | 2.15 (2) | -0.43 (3) | 66 (2) | not determined |
| 300 K | | $R_B$ (%):2.00 | $R_f$ (%):1.07 | $R_M$ (%):17.2 | $\chi^2$: 12.2 |
| AF2 $\mathbf{k_2}$= (1/2, 1/2, 1/2±*0.102*) | 175 | 2.19 (3) | -0.44 (15) | 47 (6) | - |
| 10 K | | $R_B$ (%):2.12 | $R_f$ (%):1.22 | $R_M$ (%):15.6 | $\chi^2$: 13.2 |

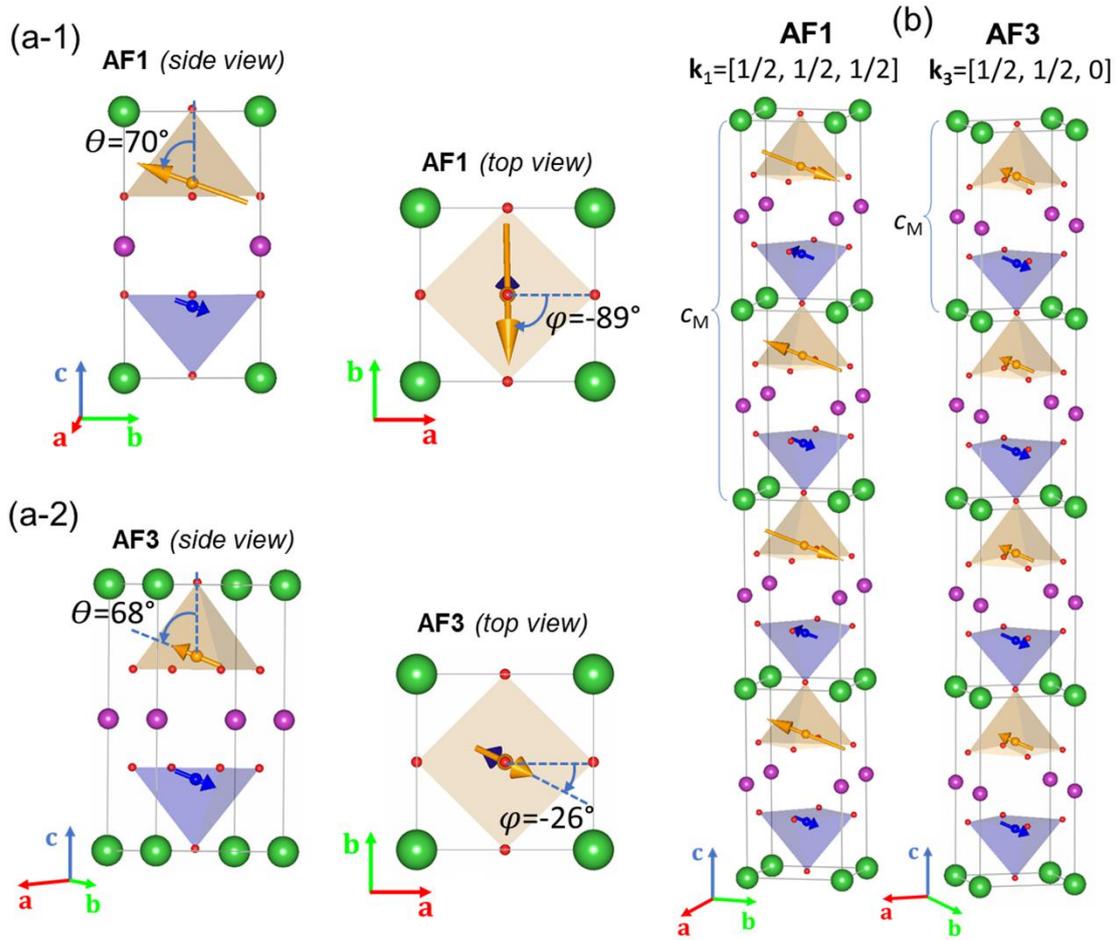

**Fig. 5. Single crystal.** (a) Two projections of the chemical unit cell illustrating the tilting ($\theta$) of the magnetic moments respect to the *c*-axis and the orientation ($\varphi$) of the $m_{\perp c}$ component in the *ab*-plane for the AF1 (a-1) and AF3 (a-2) phases. (b) Projections of the refined AF1 (left) and AF3 (right) collinear magnetic phases including four unit cells to illustrate the different translational symmetries along the *c*-direction. In the unit cell, only the majority atoms are shown in the pyramids (brown: Fe[Mn]; blue: Cu).

Notice that the moments reported are averaged over the total crystal volume, and the specific volume ratio occupied by each phase is unknown. Typically, smaller intensities were observed in AF3 magnetic reflections as compared to AF1 ones, which suggests that the former occupies a smaller fraction of the crystal. Still, for the AF1 collinear phase, the obtained average ordered magnetic moment obtained at the upper and lower pyramids of the chemical cell are rather different, namely an m(Fe)/m(Cu)=3.19 ratio is found (at 50 K). This contrasts with the moments in the AF3 phase, which are much more similar (m(Fe)/m(Cu)=1.57). This result provides strong evidence that the secondary AF3 phase appears in regions of the crystal presenting a very large Fe/Cu cation disorder. Given the high level of Fe/Cu cationic order found in this crystal, the AF3 phase very likely comes from regions with a higher presence of Mn atoms.

Regarding the orientation of the magnetic moments, several conclusions can be drawn from Table II. First, the refined tilt of the spins respect to the *c*-axis in the AF1 phase ($\theta$ =70(2)°) is closely similar to its homologue YBaCuFe$_{0.95}$Mn$_{0.05}$O$_5$ polycrystalline sample ($\theta$ =66(2)° [19]). Second, the tilt angle found for the AF3 phase ($\theta \approx$ 68(3)°) is within errors the same as in the AF1 phase. Hence, the inclination of the easy-axis is not modified when comparing the two collinear phases with different magnetic cells. Third, differences are found regarding the orientation of the moment component parallel to the *ab*-plane (m$_{\perp c}$). It is worthwhile highlighting that for the AF1 phase the refinements unambiguously uncover that the m$_{\perp c}$ component is practically parallel to the *b*-axis ($\varphi$=-89(2)°), thus breaking the tetragonal symmetry. Solutions along the diagonal direction or parallel to the *a-axis* led to clearly worse agreement factors. In the AF3 phase the m$_{\perp c}$ component was found not clearly aligned along any of the tetragonal axes, but adopts an intermediate direction ($\varphi$=-26(3)°). The tilting of the magnetic moments respect to the *c*-axis and their orientation within the *ab*-plane are exposed in Fig. 5(a) for the two coexisting phases in the crystal. Additionally, Fig. 5(b) shows a schematic projection of the refined AF1 and AF3 magnetic phases embracing four unit cells to illustrate their different translational symmetry along the *c*-direction.

Notice that for the AF3 collinear phase the magnetic anti-translation for successive cells along *c* is suppressed ($k_z$=0), meaning that the coupling between spins at the bipyramids is antiferromagnetic. This is in contrast with the AF1 order, where the spins sharing a bipyramid are ferromagnetically coupled in virtue of the ferromagnetic Fe$^{3+}$-O-Cu$^{2+}$ exchange interaction along *c*. In minority regions presenting higher chemical disorder the weak ferromagnetic Fe-O-Cu bond has been substituted by the strong AFM Fe-O-Fe pair. Much weaker is the Fe$^{3+}$-O-Mn$^{3+}$ (or Mn$^{3+}$-O-Mn$^{3+}$) AFM coupling within the bipyramids. Additional information on the spiral phase (as e.g. its evolution with temperature) in the powder sample can be found in ref. [12].

## 4. Conclusion

The structural and magnetic properties of the candidate chiral multiferroic YBaCuFe$_{0.95}$Mn$_{0.05}$O$_5$ have been studied on a single crystal specimen, in comparison with a polycrystalline sample synthesized by solid-state reaction (last cooling rate of 300 K/h). The high-quality single crystal was grown by a modified traveling solvent floating zone technique, and its characterization has included single-crystal neutron diffraction measurements down to 50 K. The study has revealed that the crystal displays very different properties as compared to the powder. First, the B-site disorder was quantified, obtaining that the Fe/Cu chemical disorder in the single crystal is ~10%, significantly lower than in the analogous powder sample where disorder is ~30%. In light of this finding, it is important to highlight that the TSFZ method used to grow the crystal produced a layered perovskite with a much higher level of Fe/Cu cationic order at the B/B' sites than the powder sample. We have shown that this fact has deep implications on the magnetic properties of the system. The most relevant one is the lack of ICM spiral phase in the single crystal specimen, whereas the powder sample adopts a spiral phase below $T_{N2} \approx 175$ K ($\mathbf{k_2}$=(1/2 1/2 1/2±q) with q=0.102 r.l.u.). The $\mathbf{k_1}$-collinear transition temperature of the crystal ($T_{N1} \approx 420$ K) was close to that of the powder (438 K). In addition, single-crystal neutron diffraction reveals that a second collinear phase with propagation vector $\mathbf{k_3}$=(1/2 1/2 0) develops in the crystal below $T_{N3} \approx 242$ K. The two collinear magnetic structures of the crystal were determined. Apart from observing the same inclination in both specimens of the AF1 magnetic easy-axis respect to *c*, it is worthwhile highlighting that the moment orientation in the *ab*-plane was found parallel to the *b*-axis in the crystal, discarding the diagonal direction. The second collinear phase comes from regions in the crystal with higher cationic disorder where the Fe-Cu ferromagnetic coupling between spins sharing a bipyramid becomes antiferromagnetic due to disorder.

These results give strong support to the "*order by disorder*" theoretical models intended to justify the high-temperature stability of the spiral phase in these compounds. They equally demonstrate the strong influence of the preparation method in obtaining the potential chiral and multiferroic properties in this family of perovskites.

**Acknowledgements.**

This work has received financial support from the Spanish Ministerio de Ciencia, Innovación y Universidades (MINCIU), through Projects No. RTI2018-098537-B-C21 and RTI2018-098537-B-C22, cofunded by ERDF from EU, and "Severo Ochoa" Programme for Centres of Excellence in R&D (FUNFUTURE (CEX2019-000917-S)). X.Z. was financially supported by China Scholarship Council (CSC) with No. 201706080017. X.Z's work was done as a part of the Ph.D program in Materials Science at Universitat Autònoma de Barcelona. We also acknowledge ALBA, ILL and D1B-CRG (MINECO) for provision of beam time (dois: 10.5291/ILL-DATA.CRG-2655, 10.5291/ILL-DATA.CRG-2478, 10.5291/ILL-DATA.CRG-2562).